\documentclass[preprint2]{aastex}
\usepackage{appendix}
\usepackage{color}
\usepackage{amsmath}

\begin{document}

\title{Interpreting the relation between the gamma-ray  and infrared luminosities of star-forming galaxies}

\author{Yi Zhang\altaffilmark{1,2}, Fang-Kun Peng\altaffilmark{3,4}, Xiang-Yu Wang\altaffilmark{1,2}}

\altaffiltext{1}{School of Astronomy and Space Science, Nanjing University, Nanjing 210023, China;
xywang@nju.edu.cn}
\altaffiltext{2}{Key laboratory of Modern Astronomy and Astrophysics (Nanjing University), Ministry of Education, Nanjing 210023, China}
\altaffiltext{3}{School of Physics and Electronic Science, Guizhou Normal University, Guiyang 550001, China; pengphy@gznu.edu.cn}
\altaffiltext{4}{Guizhou Provincial Key Laboratory of Radio Astronomy and Data Processing, Guizhou Normal University, Guiyang 550001, China}

\begin{abstract}
It has been found that there is a quasi-linear scaling relationship between the gamma-ray luminosity in GeV energies and the total infrared luminosity of star-forming galaxies, i.e. $L_{\gamma}\propto L_{\rm IR}^{\alpha}$ with $\alpha\simeq 1$. However, the origin of this linear slope is not well understood. Although extreme starburst galaxies can be regarded as calorimeters for hadronic cosmic ray interaction and thus a quasi-linear scaling may hold, it may  not be the case for  low star-formation-rate (SFR) galaxies, as the majority of cosmic rays in these galaxies are expected to escape. We calculate the gamma-ray production efficiency in star-forming galaxies by considering realistic galaxy properties, such as the gas density and galactic wind velocity in star-forming galaxies. We find that the slope for the relation between gamma-ray luminosity and the infrared luminosity gets steeper for low infrared luminosity galaxies, i.e. $\alpha\rightarrow 1.6$, due to increasingly lower efficiency for the production of  gamma-ray emission. We further find that the measured data of the gamma-ray luminosity is compatible with such a steepening. The steepening in the slope suggests that cosmic-ray escape  is very important in low-SFR galaxies.
\end{abstract}


%%%%%%%%%%%%%%%%%%%%%%%%%%%%%%%%%%%%%%%%%%%%%%%%%%%%%%%%%%%%%%%%
\section{Introduction}
Nearby star-forming and starburst galaxies have been identified to be GeV-TeV gamma-ray sources \citep{2009Sci...326.1080A,2009Natur.462..770V,2010ApJ...709L.152A,2012ApJ...755..164A}. Cosmic rays (CRs) accelerated by supernova remnants or stellar winds interact with the interstellar medium (ISM) and produce neutral pions (schematically written as $p+p\rightarrow \pi^0$+other products), which in turn decay into high-energy gamma-rays ($\pi^0\rightarrow \gamma+\gamma$).
Using the two-year observations obtained with the Large Area Telescope (LAT) aboard the \textsl{Fermi} mission, \citet{2010A&A...523L...2A} found a correlation between the gamma-ray luminosity and star-formation-rate (SFR) for the Local Group galaxies.
With a larger galaxy sample, \citet{2012ApJ...755..164A} found a quasi-linear scaling relation between gamma-ray luminosity and infrared luminosity which applies to both  quiescent galaxies of the Local Group and low-redshift starburst galaxies. Since young stars in star-forming galaxies emit ultraviolet emission which  is absorbed by dust in the ISM and reprocessed into infrared (IR) emission, the IR luminosity is a good indicator of SFR. Recently, the scaling relation was extended to an even  higher IR luminosity regime, with detection of GeV emission from nearby luminous and ultraluminous infrared galaxies \citep{2014ApJ...794...26T,2016ApJ...821L..20P,2016ApJ...823L..17G}.

The quasi-linear correlation between the gamma-ray luminosity $L_{\gamma}$ in the energy range $0.1-100\ \rm  GeV$ and the total infrared luminosity $L_{\rm IR}$ ($8-1000\ \mu m$) has the form \citep{2012ApJ...755..164A}
\begin{equation}
{\rm log}\left(\frac{L_{\gamma}}{\rm erg\ s^{-1}}\right)=\alpha {\rm log}\left(\frac{L_{\rm IR}}{10^{10}L_\odot}\right)+\beta,
\end{equation}
where a nearly linear power-law index of $\alpha=1.0-1.2$ is obtained \citep{2012ApJ...755..164A}.
Since the infrared luminosity reflects the cosmic ray injection rate in the galaxy, the quasi-linear relationship implies that a constant fraction of cosmic ray energy is converted into gamma-rays across all these galaxies.
While, theoretically, extreme starburst galaxies, such as Arp 220, can be considered as cosmic ray calorimeter (e.g., \citep{2004ApJ...617..966T,2011ApJ...734..107L,2015MNRAS.453..222Y,2018MNRAS.474.4073W}), i.e., almost all of the cosmic ray energy is converted into secondary particles due to high gas density, CRs in low-SFR  galaxies are expected to transfer only a small fraction of energy into secondary particles due to a much lower gas density in these galaxies.
{For  starburst galaxies (SBGs) such as M82, the optical depth for $pp$ interactions over typical SBG size is low and CRs escape mostly unscathed, e.g., for a high average gas density, $n_ {\rm H} = 10^2 \ \rm cm^{-3}$, the mean free path of a few TeV proton is $65 \ \rm kpc$, much larger than the typical disk radius (e.g., \cite{2008A&A...486..143P}). A consequence is that these SBGs  are inefficient calorimeters (at $\sim 30\%$ level, see \cite{2011ApJ...734..107L}), and normal star-forming galaxies have even lower fractions of cosmic-ray energy  converted into gamma-rays.}
Thus, the origin of this  linear slope is puzzling.

The aim of this paper is to interpret this apparent quasi-linear relationship. We first calculate the gamma-ray production efficiency in star-forming galaxies by considering realistic galaxy properties, such as the gas density and galactic wind velocity in star-forming galaxies. It is shown that the efficiency of producing gamma-rays becomes increasingly low for low infrared luminosity galaxies. We further find that this theoretic model can reproduce the observed relation between the gamma-ray luminosity and infrared luminosity. Motivated by such a gradually changing efficiency, we also use an empirical function, i.e., a smoothly broken power-law function, to fit the observation data, and find that the scaling index at the low luminosity end deviates significantly from the linear relation, which is consistent with our theoretical expectation.

The rest of this paper is structured as follows. In Section 2, we describe the formulas  of calculating the gamma-ray emission related to IR luminosity. In section 3, we present the results. We give our discussions in Section 4 and conclusions in Section 5.

\section{Gamma-ray production in star-forming galaxies}
Massive stars in galaxies end their lives as core-collapse supernovae, whose remnants can accelerate CRs and inject them into the ISM.
CRs produce high-energy gamma rays through inelastic collisions  with ISM  ($pp$ collisions), electron and positron bremsstrahlung, and inverse Compton (IC) scattering of the primary and secondary electrons.
Detailed calculations have shown that the pionic decay gamma-rays dominate the emission above $100\ \rm MeV$ for star-froming galaxies (e.g., \cite{2005A&A...444..403D,2010MNRAS.401..473R}), although leptonic emission is expected to become increasingly important at lower energies.

Therefore, we expect the  injection power of CRs is proportional to SFR, i.e.,  $L_{\rm CR}\propto {\rm SFR}$. The gamma-ray luminosity produced by $pp$ collisions between CRs and ISM can be parameterized by $L_\gamma \propto f_\pi \, L_{\rm CR}$ , where $f_\pi$ is the efficiency of CR energy transferred to secondary pions. On the other hand, CRs can escape from galaxy through diffusion and galactic wind advection. These two processes compete to generate the efficiency
\begin{equation}
f_\pi = 1-{\rm exp}(-\frac{t_{\rm esc}}{t_{\rm loss}}),
\end{equation}
where $ t_{\rm esc}$ is the escape time of CRs and $t_{\rm loss}$ is the energy loss time of CRs via $pp$ collisions.

The CR energy loss timescale depends on the average gas density and the inelastic collision cross section, given by $t_{\rm loss}=(0.65 n \sigma_{pp} c)^{-1}$  \citep{1994A&A...286..983M}, where $\sigma_{pp}$ is the inelastic cross section. For CR energy larger than $1 \ \rm GeV$, $\sigma_{pp}$ is nearly a constant with energy, and we take the value $2.5\times10^{-26}\ \rm cm^2$.  The ISM number density $n$ relates with the gas surface density $\Sigma_g$  by $\Sigma_g = n m_p  H $, where $m_p$ is mass of proton, $H$ is the scale height of galaxy disk. Then the CR energy loss time is given by
\begin{equation}
t_{\rm loss}=3.36\times 10^5 \frac{H}{1\,{\rm kpc}} \left(\frac{\Sigma_g}{1\,{\rm g\,cm^{-2}}}\right)^{-1}\ {\rm yr}.
\end{equation}

CRs escape out of the galaxy in the form of diffusion or galactic wind advection.
In the case of diffuse process, CRs are scattered by small-scale inhomogeneous magnetic fields. Diffusive time scales can be approximated as $t_{\rm diff}=H^2/{4D}$, where $D=D_0\,(E/E_0)^\delta$ is the diffusion coefficient, $D_0$ and $E_0$ are normalization factors. We take the  standard diffusion coefficient for ISM as $D=3.86\times10^{28}(E_p/{\rm GeV})^{1/3}\ {\rm cm^2 \ s^{-1}}$, where the value of $\delta=1/3$ is for the Kolmogorov type turbulence. Then the diffusion time is given by
\begin{equation}
t_{\rm diff}=1.97\times 10^6 \left(\frac{H}{\rm kpc}\right)^2 \left(\frac{E_p}{\rm GeV}\right)^{-\frac{1}{3}}\ {\rm yr} \label{2}.
\end{equation}
In the case of advection process, CRs are transported outward with galactic wind on a characteristic timescale
\begin{equation}
t_{\rm adv}=\frac{H}{v_w}=1.96\times 10^6 \frac{H}{1\ {\rm kpc}}\left(\frac{v_w}{500{\rm \ km\ s}^{-1}}\right)^{-1}\ {\rm yr},
\end{equation}
where $v_w$ is the velocity of the galactic wind.
The timescale for CRs escaping out of the galaxy is parameterized as
\begin{equation}
t_{\rm esc}=\frac{1}{t_{\rm diff}^{-1}+t_{\rm adv}^{-1}}.
\end{equation}

Next, we calculate the pion production efficiency by considering realistic galaxy properties, such as  gas surface density $\Sigma_g$, galactic wind velocity $v_w$ and galaxy scale height $H$. The total IR luminosity in 8-1000$\mu m$ is one well-established tracer of the SFR for late-type galaxies, so we take \citep{1998ApJ...498..541K}
\begin{equation}
\frac{{\rm SFR}}{M_\odot {\ \rm yr}^{-1}}=\epsilon\,1.7\times 10^{-10}\frac{L_{\rm IR}}{L_\odot}
\end{equation}
where the factor $\epsilon=0.79$ is for initial mass function (IMF) derived from \citet{2003ApJ...586L.133C}, and $\epsilon=1$ is for initial mass function used by \cite{1998ApJ...498..541K}. Below we use the IR luminosity instead of SFR in the subsequent calculation.

To determine the gas surface density $\Sigma_g$, we use the classical Kennicutt-Schmidt law (K-S law) that extends over several orders of magnitude in SFR and gas density \citep{1998ApJ...498..541K}:
\begin{equation}
\begin{split}
\Sigma_{\rm SFR}=(2.5 \pm0.7)\times10^{-4} \left(\frac{\Sigma_g}{M_\odot\ {\rm pc}^{-2}}\right)^{1.4\pm0.15}\\ M_\odot \ {\rm yr^{-1} \ kpc}^{-2}
\end{split}
\end{equation}
where $\Sigma_{\rm SFR}$ is disk-averaged SFR density. $\Sigma_{\rm SFR}$ can be obtained from $\Sigma_{\rm SFR}={\rm SFR}/\pi R^2$. Replacing SFR with  the IR luminosity, K-S law can be written as
\begin{equation}
\begin{split}
\Sigma_g=(10.43\pm 2.41)\times10^{-5} \left(\frac{L_{\rm IR}}{L_\odot}\right)^{\frac{1}{1.4\pm0.15}} \\ \left(\frac{R}{\rm pc}\right)^{-\frac{2}{1.4\pm0.15}} {\rm g\ cm}^{-2},
\end{split}
\end{equation}
where $\epsilon=1$ for the IMF used by \cite{1998ApJ...498..541K} is adopted. Galactic winds or gaseous outflows from  starburst galaxies can accelerate CRs to high speed. For luminous infrared
galaxies at low redshift, winds from more luminous starbursts
have higher speeds roughly as $v_w\propto {\rm SFR}^{0.35}$ \citep{2005ApJ...621..227M}. A
similar relation is found for star-forming galaxies at $z = 1.4$ \citep{2009AIPC.1201..142W}.  Thus, we have
\begin{equation}
\begin{split}
v_w & =10^{1.56\pm0.13} \left(\frac{\rm SFR}{M_
 \odot\ {\rm yr}^{-1}} \right)^{0.35\pm0.06} \\
& =10^{1.56\pm0.13} \left(\frac{L_{\rm IR}}{5.8\times10^9\,L_\odot} \right)^{0.35\pm0.06} {\rm km\ s}^{-1}.
\end{split}
\end{equation}

Considering that the relation between galaxy scale height and SFR is not  straightforward, we assume galaxy height relates with galaxy radius as $H=(0.21\pm 0.02)\ R$ \citep{2008MNRAS.388.1321P}, where $R$ is the disk radius of a galaxy.  For late-type galaxies, there is a relation between the radius of galaxy and the total stellar mass, given by \citep{2003MNRAS.343..978S}
\begin{equation}
\bar{R}=0.1\left(\frac{1.26M_*}{M_\odot}\right)^{0.14}\left(1+\frac{1.26M_*}{3.98\times10^{10}\,M_\odot}\right)^{0.25} {\rm kpc},
\end{equation}
with the dispersion of
\begin{equation}
\sigma_{\ln_R}=0.34+\frac{0.13}{1+\left(\frac{1.26M_*}{M_\odot}\right)^2} {\ \rm kpc}.
\end{equation}
For star-forming galaxies in local universe, based on the tight relationship between total stellar mass and SFR of galaxy \citep{2010ApJ...721..193P}, we get  ${\rm SFR}=7.94(\rm M/{10^{11}\rm M_\odot})^{0.9}\ \rm M_\odot \ {\rm yr}^{-1}$, where the dispersion is  0.3 dex. The SFR used in \cite{2010ApJ...721..193P} is computed for the Kroupa IMF, so we convert it to the case for the Chabrier IMF by using ${\rm log}({\rm SFR}_{\rm Chabrier})={\rm log}({\rm SFR}_{\rm Kroupa})-0.04$. Then the total stellar mass relates with the IR luminosity by
\begin{equation}
M_*=10^{11}\,\left(\frac{L_{\rm IR}}{5.38_{-2.69}^{+5.38}\times10^{10}L_\odot}\right)^{1.11}.
\end{equation}

Using the above relations, we take galaxy parameters, such as gas surface density $\Sigma_g$ , galactic wind velocity $v_w$ and galaxy scale height $H$, as  input parameters and generate the pion production efficiency $f_\pi$ as an output, which is in turn  used to  produce the $L_{\gamma}-L_{\rm IR}$ relation
\begin{equation}
L_\gamma = L_{\pi _ 0}=C f_\pi {L_{\rm IR}},
\end{equation}
where $L_\gamma$ is the total energy of gamma-rays as integrating over corresponding energy range, $L_{\pi _ 0}$ is the total energy of $\pi_0$ produced by $pp$ collisions and $C$ is a normalization factor. The normalization factor $C$ can be obtained using the data of M82.

\section{Results}

\begin{figure}
\centering
\includegraphics[scale=0.5]{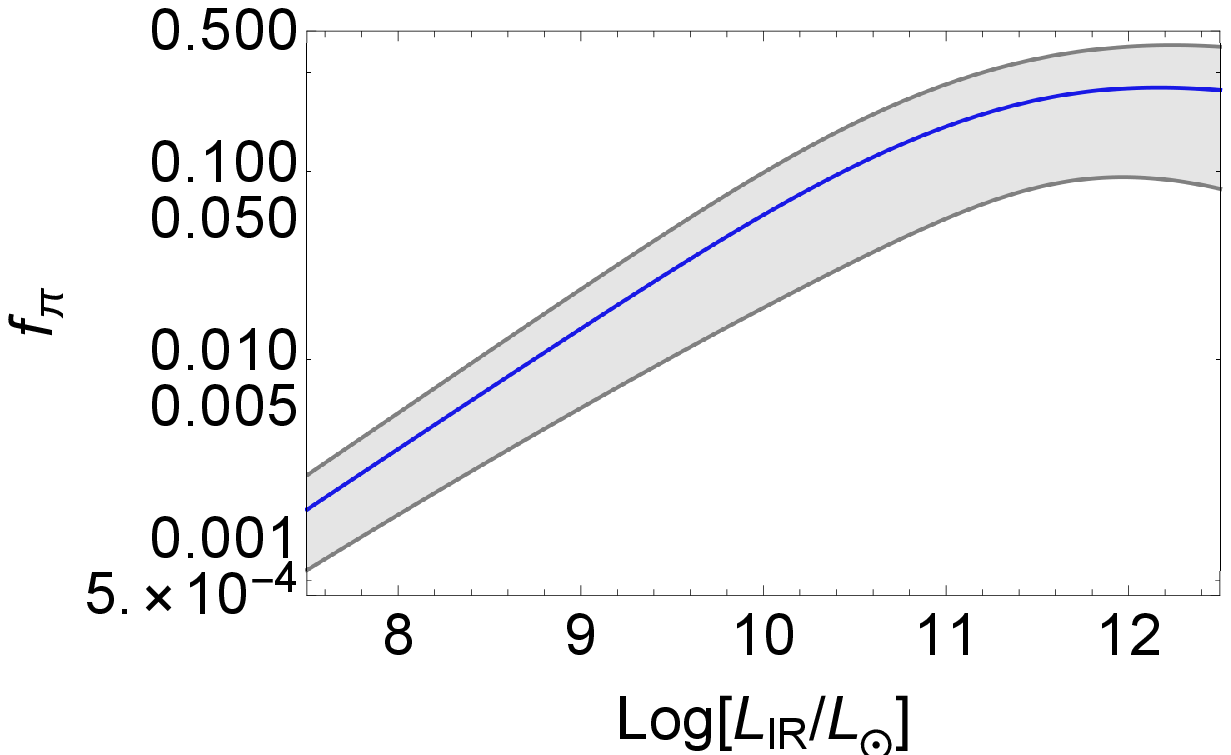}
\caption{Pion production efficiency $f_\pi$ versus IR luminosity ($L_{\rm IR}$) for star-forming galaxies. The shaded region denotes the uncertainty of $f_\pi$ resulting from uncertainties in the relations among galaxy parameters. }
\label{figpi}
\end{figure}

The pion production efficiency $f_\pi$ is shown in Fig. \ref{figpi}.
The uncertainty given in Fig. \ref{figpi} takes into account of all the uncertainties in the relation between the specific galaxy parameter and IR luminosity. One can see that the $pp$ interaction is quite inefficient in low IR luminosity galaxies, with efficiency being less than $5\%$ for $L_{\rm IR} < 10^{8} L_{\odot}$. For galaxies with higher IR luminosity  $>10^{11}L_\odot$, the efficiency increases to about $25\%$, albeit with a somewhat larger uncertainty.

Now we compare our theoretical prediction with the observed gamma-ray luminosity of star-forming galaxies  (Fig. \ref{figLumIR}). Considering that gamma-rays at $100 \ \rm MeV$ may be contaminated by leptonic emission or point sources like pulsars, we use a higher threshold energy of $1 \ \rm GeV$ for our study. It is believed that $1-500\ \rm GeV$ gamma-ray emission of star-forming galaxies and star-burst galaxies has a hadronic origin. The parameters of the GeV-detected star-forming galaxies, including $L_{1-500\ \rm GeV }$, $L_{\rm IR}$ and SFR, are listed in Table \ref{table1}. The data of $L_{1-500\ \rm GeV }$ are taken from \cite{2019A&A...621A..70P}.
From  Fig. \ref{figLumIR}, one can see that the {theoretical} model for the gamma-ray luminosity ($L_{1-500 \ \rm GeV}$) (the blue lines) {is in compatible (within the low available statistics) with} observations, especially for galaxies with  IR luminosity $L_{\rm IR}$ above $10^9 L_\odot$. The slope for the relation between the gamma-ray luminosity and IR luminosity is not a constant. For galaxies with higher IR luminosity, the slope of the curve approaches to $1$, but it steepens to $\sim 1.6$ for low infrared luminosity galaxies, which reflects an increasingly lower efficiency for the production of  gamma-ray emission (see Fig. \ref{figpi}).
This means that high luminosity galaxies are {closer to (but still short of) being} CR calorimeters (shown by the gray line), while low luminosity galaxies deviate from the calorimetric limit significantly. {The model can explain the data obtained in Ackermann et al. (2012) and Peng et al. (2016).}
We will discuss this in more detail in the next section.

\section{Discussions}
\begin{figure}
\centering
\includegraphics[scale=0.6,clip]{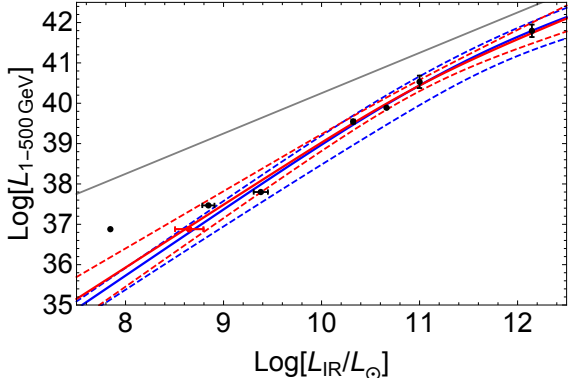}
\caption{The derived gamma-ray luminosity versus IR luminosity of star-forming galaxies. The solid blue line denotes the derived gamma-ray luminosity using our model.
The solid red line denotes the best-fit using a smoothly broken power-law function.
The corresponding dashed lines indicate the  uncertainties at $1 \ \sigma$ confidence level.
The gray line denotes gamma-ray luminosity in the calorimetric limit (i.e., $f_\pi=1$). The black dots represent the GeV-detected galaxies listed in Table 1.
The red dot represents the Small Magellanic Cloud (SMC) with a corrected IR luminosity (see the text for the details).}
\label{figLumIR}
\end{figure}

As shown in Fig \ref{figLumIR}, the Small Magellanic Cloud (SMC) is an apparent outlier of the correlation. We suggest that this could be due to the underestimate of SFR for SMC.
The IR luminosity can be regarded as a well-established tracer of SFR  only when the IR emission of interstellar dust is nearly calorimetric measure of radiation produced by young stellar populations \citep{2003ApJ...586..794B}.
That is, for galaxies with IR luminosity larger than $10^9L_\odot$, the IR luminosity is a robust SFR indicator, but for low IR luminosity galaxies such as SMC, the substantially low metallicity and dust content lead to a low optical depth for IR photons. As a result,  the observed IR emission only reflects a fraction of the star formation activity, and thus the SFR of these galaxies are underestimated.
After considering the combined observations of $ H_\alpha$ and IR emission, we estimate that the SFR of SMC is $0.04-0.08\ M_\odot\ {\rm yr}^{-1}$ \citep{2004A&A...414...69W}.
Using  Eq.(7) to convert the SFR into IR luminosity, we find that the IR luminosity of SMC is corrected to $\sim 4.5\times10^8 L_\odot$.
The new data for SMC is shown by a red dot in Fig. \ref{figLumIR} and one can see that it agrees well with our theoretical model.

\begin{figure}
\centering
\includegraphics[scale=0.6,clip]{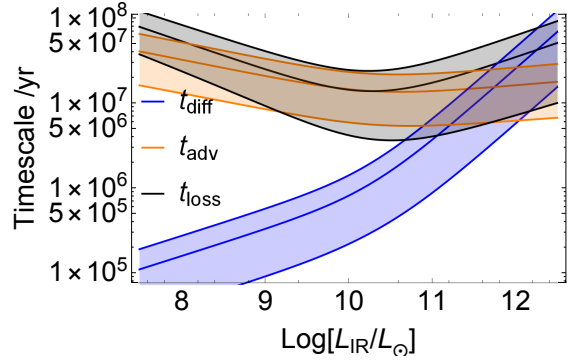}
\caption{The timescales of three processes for CRs in star-forming galaxies: $pp$ collision, diffusion process and advection transport. The shaded regions denote the uncertainty accordingly.}
\label{figtime}
\end{figure}

The finding of a changing slope for the relation between the gamma-ray luminosity and the total IR luminosity suggests that the mechanism dominating the CR energy loss process is different for different IR-luminosity galaxies.
We now discuss the physical process accounting for the scaling slope change.
The CR energy loss timescale through $pp$ collision and escape timescales through diffusion and advection are shown in Fig. \ref{figtime}.
One can see that the diffusion escape is the fastest when the IR luminosity is less than $10^{11}L_\odot$, leading to a low efficiency of pion production and a steep slope.
This situation changes when the IR luminosity is larger than $10^{11}L_\odot$. The diffusion timescale increases sharply due to the increase of the galaxy disk scale height.
The advection escape becomes increasingly important  in more luminous star-forming galaxies due to higher galactic wind velocities.
For the timescale of $pp$ collision, it declines first due to increases of the gas density and then rise due to the relatively larger galaxy disk scale height.
The timescales of the advection and the energy loss through $pp$ collisions are on the same order of magnitude, thus CRs  collide with ISM  effectively before escaping out of the galaxy.

Since the derived relation between $L_{1-500 \ \rm GeV}$ and $L_{\rm IR}$ indicates a gradually changing slope, we try to use a smoothly broken power law (SBPL) function to fit the data. For more details about the SBPL fit, please refer to the Appendix.
As it is difficult to constrain all the parameters of  the SBPL model for a small number of the sample galaxies, we fix the slope at the high IR-luminosity end as $\alpha=1$, which is motivated by the calorimetric limit for these galaxies.
We further remove SMC from the sample as its IR luminosity is not an accurate SFR tracer.
We use the maximum likelihood approach as illustrated in our previous work \citep{2019A&A...621A..70P}.
We find that the slope $\alpha$ for low IR-luminosity galaxies approaches to $1.56\pm0.15$  as the IR luminosity decreases.
The result is shown in Fig. \ref{figLumIR} by the red lines. The  slope found for this empirical fit function  is well consistent with our theoretical model.

\begin{figure}
\centering
\includegraphics[scale=0.6,clip]{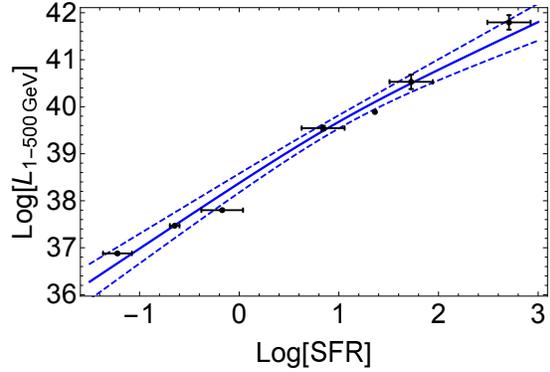}
\caption{The fit of the relation between gamma-ray luminosity and SFR with a smoothly broken power-law function. The dashed lines indicate the uncertainty at $1 \ \sigma$ confidence level.}
\label{figfit}
\end{figure}

As mentioned above, the IR luminosity underestimates SFR for low IR-luminosity galaxies. To eliminate this issue,  we study the relation using SFR directly.
The result is shown in Fig. \ref{figfit}.
The slope changes from $1$ at high SFR end to $1.41\pm0.15$ at low SFR end for a fit with a smoothly broken power-law function.
We compare the maximum likelihood of the two fits with a smoothly broken power-law function and a single power-law function, and find that the difference in the  maximum likelihood between the two
models is not significant ($< 2$). From this point of view, we think that the smoothly broken power law function is equally acceptable by the data.

\section{Conclusions}
We calculate the gamma-ray luminosity of star-forming galaxies by considering realistic galaxy properties, such as gas surface density, galaxy scale height and wind velocity.
The derived relation between the gamma-ray luminosity and IR luminosity shows a gradually changing slope, as expected from increasingly lower efficiency of gamma-ray production for low luminosity galaxy. We further find that the measured data  is well consistent with such a changing slope. As an comparison, we use a smoothly broken power law function to fit the data and find that the slope for low IR luminosity galaxies deviates from the linear  relation significantly, but agrees well with our theoretical calculation. Our result suggests that most CRs escape before significant collisional losses in low-SFR galaxies, mainly through the diffusion process. This is consistent with the findings in some recent numerical simulations \citep{2017ApJ...847L..13P,2018arXiv181210496C}.

\section*{Acknowledgments}
X.Y.W. is supported by the National Key R \& D program of China under the grant 2018YFA0404203 and the NSFC  grants
11625312 and 11851304. F.K.P acknowledges support from the Doctoral Starting up Foundation of Guizhou Normal University 2017 (GZNUD[2017] 33).

\clearpage

\begin{table}
\centering
\caption{ Parameters of GeV-detected star-forming galaxies}
\begin{tabular}{lccc}
\hline
\hline
Name&$L_{1-500 \ {\rm GeV}}$ &$L_{\rm IR}$&SFR\\
 &$\rm erg\ s^{-1}$&$10^9 L_\odot $&$M_\odot\ \rm yr^{-1}$\\
\hline
SMC&$(7.57\pm0.34)\times10^{36}$&$0.07\pm0.01$&0.04-0.08\tablenotemark{a}\\
LMC&$(2.94\pm0.05)\times10^{37}$&$0.7\pm0.1$&0.20-0.25\tablenotemark{b}\\
M31&$(6.31\pm0.21)\times10^{37}$&$2.4\pm0.4$&0.35-1\tablenotemark{c}\\
NGC 253&$(3.51\pm0.40)\times10^{39}$&21&3.5-10.4\tablenotemark{d}\\
M82&$(7.74\pm0.28)\times10^{39}$&46&13-33\tablenotemark{e}\\
NGC 2146&$(3.39\pm1.20)\times10^{40}$&100&26.6-79.7\tablenotemark{f}\\
Arp 220&$(6.24\pm2.21)\times10^{41}$&1400&254.8-764.3\tablenotemark{f}\\
\hline
\end{tabular}

\textbf{Notes.} The $1-500\ {\rm GeV}$ gamma-ray luminosities are taken from \cite{2019A&A...621A..70P}, IR luminosities are taken from \cite{2004ApJ...606..271G}.

\textbf{References.} $^{a}$ \citet{2004A&A...414...69W}; $^{b}$\citet{2007MNRAS.382..543H}; $^{c}$\citet{2009A&A...505..497Y}; $^{d}$\citet{2006AJ....132.1333L};  $^{e}$\citet{2003ApJ...599..193F};  $^{f}$\citet{2005ApJ...621..139C}.
\label{table1}
\end{table}

\clearpage

\appendix
The smoothly broken power law functon adopted here is in the form of \footnote{\url{http://docs.astropy.org/en/stable/api/astropy.modeling.powerlaws.SmoothlyBrokenPowerLaw1D.html}}
\begin{eqnarray}
f(x)=A\left(\frac{x}{x_b}\right)^{\alpha_1}\left[\frac{1}{2}+\frac{1}{2}\left(\frac{x}{x_b}\right)^{1/\delta}\right]^{(\alpha_2-\alpha_1)\delta},
\end{eqnarray}
where $\alpha_1$ is the slope for $x \leq x_b$ and $\alpha_2$ for $x\geq x_b$, $\delta$ is smoothness parameter denoting the degree of slope change.
The two power laws are smoothly joined at values $x_1 < x< x_2$.
The change of slope occurs between the values $x_1$ and $x_2$ such that:
\begin{equation}
\delta=\log_{10}\frac{x_2}{x_b}=\log_{10}\frac{x_b}{x_1}.
\end{equation}

We apply the above function to study the correlation analysis of our sample star-forming galaxies, using the following forms:
\begin{equation}
\log_{10}L_{1-500\ {\rm GeV}}=A+\alpha\,(\log_{10}L_{\rm IR}-\log_{10}L_b)- \delta\,(\alpha_1-\alpha_2)\times\log_{10}\left[\frac{1}{2}+\frac{1}{2}\left(\frac{L_{\rm IR}}{L_b}\right)^{1/\delta}\right].
\end{equation}
Due to a small galaxy sample, we fix the break IR luminosity at $L_{\rm b}=10^{11}\ L_\odot$ and take $\delta=1$. We also fix the index at high IR luminosity as $\alpha_2 =1$, as the calorimetric limit predicts.

In the case studying the relation between the gamma-ray luminosity and SFR, we use the form
\begin{equation}
\log_{10}L_{1-500\ {\rm GeV}}=A+\alpha\,(\log_{10} {\rm SFR}-\log_{10}{\rm SFR}_b)- \delta\,(\alpha_1-\alpha_2)\times\log_{10}\left[\frac{1}{2}+\frac{1}{2}\left(\frac{\rm SFR}{{\rm SFR}_b}\right)^{1/\delta}\right].
\end{equation}
Similarly, we take the break point at ${\rm SFR}=10M_\odot\ {\rm yr}^{-1}$,  $\delta=1$, and fix the index at the high-SFR end as $\alpha_2 =1$.

\end{document}